\documentclass[12pt]{article}
\usepackage{amsmath, amssymb, amsfonts}
\usepackage[T1]{fontenc}
\usepackage[latin1]{inputenc}
\usepackage{authblk}
\usepackage{tikz}
\usetikzlibrary{shapes,arrows}
\usepackage{graphics}
\usepackage{verbatim}
\usepackage{color}
\usepackage{hyperref}

\begin{document}

\title{All the Groups of Signal Analysis\\from the  $(1+1)$-affine Galilei Group}
\author{S. Hasibul Hassan Chowdhury$^\dag$ \\
  and S. Twareque Ali$^{\dag\dag}$\\
\medskip
\small{Department of Mathematics and Statistics,\\ Concordia University, Montr\'eal, Qu\'ebec, Canada H3G 1M8}\\

{\footnotesize $^\dag$e-mail: schowdhury@mathstat.concordia.ca\\
 $^{\dag\dag}$e-mail: stali@mathstat.concordia.ca}}

\date{\today}

\maketitle

\begin{abstract}
We study the relationship between  the $(1+1)$-affine Galilei group and four groups of interest in signal analysis and image processing, viz., the wavelet or the affine group of the line, the Weyl-Heisenberg, the shearlet and the Stockwell groups. We show how all these groups can be obtained either
directly as subgroups, or as subgroups of central extensions of the affine Galilei group. We also study this at the level of unitary representations of the groups on Hilbert spaces.
\end{abstract}

\section{Introduction}\label{sec:intro}
There are a number of groups that are used in the current literature, on signal analysis and image processing, to construct signal transforms, as functions representing the signals over convenient parameter spaces. Of these, the most commonly used are
the wavelet group, i.e., the affine group of the real line $\mathbb R$, the Heisenberg and the Weyl-Heisenberg groups and the more recently introduced Stockwell and shearlet  groups. Another set of groups, which are extensions of the Heisenberg group by one-parameter dilations, were introduced in \cite{Taylor}. These include the shearlet group as a special case and hence are also relevant for constructing signal transforms.  As the name suggests, the wavelet group \cite{coherent,daub,torres} is used to build the well-known continuous wavelet transform while the shearlet transform, using the shearlet group \cite{Shearlet}, is applicable to situations where the signal to be analyzed has undergone shearing transformations. The Weyl or Weyl-Heisenberg group
leads to the windowed Fourier transform, useful in time-frequency analysis \cite{coherent,antetal,daub}, while the Stockwell transform \cite{stockwell1,stockwell,stockwellpap} combines features of both the wavelet and time-frequency transforms. As an interesting result, we show that the Stockwell group is just a trivial central extension of the wavelet group. (Of course, the wavelet group has  no non-trivial central extensions.) This also has the implication that the unitary irreducible representation of the Stockwell group is square-integrable over a homogeneous space (the space consisting  of the affine group parameters), a fact studied in \cite{stockwell}. 

The matrix representations of these various groups are as follows. A generic element of the Heisenberg group is given by a $3\times 3$ matrix,
\begin{equation}
  g = \begin{pmatrix} 1 & x & y \\ 0 & 1 & z \\ 0 & 0 & 1 \end{pmatrix}\; ,\qquad
  x,y,z \in \mathbb R \; ,
\end{equation}
while its one-parameter family of extensions obtained in \cite{Taylor} have the form
\begin{equation}
g =\begin{bmatrix}e^{\sigma}&ve^{\frac{\sigma}{p+1}}
&a\\0&e^{\frac{\sigma}{p+1}}&b\\0&0&1\end{bmatrix}\; , \qquad -1<p\leq 1\;,\qquad
a,b, v, \sigma \in \mathbb R \; .
\end{equation}
with the shearlet group, which is a special case ($p = 1$), being of the type,
\begin{equation}
g =\begin{bmatrix}\mu&\nu\sqrt{\mu}&\alpha\\0&\sqrt{\mu}&\beta\\0&0&1\end{bmatrix} \; , \qquad
\mu > 0,\; \nu , \alpha , \beta \in \mathbb R \;.
\end{equation}
The connected affine or wavelet group is given by $2\times 2$ matrices of the form
\begin{equation}
g =\begin{bmatrix}d&t\\0&1\end{bmatrix}\; ,\qquad
d > 0\; , t\in \mathbb R \; ,
\end{equation}
and finally, the  Stockwell group can be represented by a $4\times 4$ matrix,
\begin{equation}
g = \begin{bmatrix}1&\gamma\delta&0&\theta\\0&\gamma&0&1-\gamma\\
0&0&\frac{1}{\gamma}&0\\0&0&0&1\end{bmatrix} \; , \qquad
\gamma > 0, \; \delta , \theta \in \mathbb R \; .
\end{equation}
The question naturally arises as to whether there exists a matrix group which contains all the above groups as subgroups. It is also noteworthy that all these groups consist of upper triangular matrices.

The purpose of this paper is firstly, to answer the above question., i.e., we show how all these groups can be obtained as subgroups of various extensions of the Galilei group in $(1+1)$-dimensions. This group is a physical kinematical group,
which incorporates the symmetry of non-relativistic motion in a $(1+1)$-dimensional space-time. More precisely, we shall first extend this group by space and time dilations to obtain the  $(1+1)$-affine Galilei group, which will then be shown to contain all the above groups as subgroups, except the Stockwell group. This last group which, as we mentioned earlier,  is a trivial central extension of the wavelet group, will be obtained as a subgroup of a trivial central extension of the  Galilei-Schr\"odinger group, which itself is a subgroup of the affine Galilei group. As a second and related problem we study how unitary irreducible representations of the affine Galilei and the various centrally extended Galilei-Schr\"odinger group decompose when restricted to the above subgroups. This would shed light on how signal transforms related to the bigger groups decompose into linear combinations of transforms based on the smaller subgroups. Physically this could correspond to situations where certain parameters of a more detailed transform are averaged over or ignored.

\section{Extension to the affine Galilei group}
\label{sec:aff-gal-grp}
We start with the  (1+1)-Galilei group $\mathcal{G}_{0}$ which, as we said, is the kinematical group
of a non-relativistic space-time of $(1+1)$-dimensions. This is a three parameter group, an element of which we shall denote by $(b,a,v)$. The parameters $b$, $a$, and $v$ stand for time translation, space translation and the Galilean or velocity boost, respectively. Under the action of this group, a space-time point $(x,t)$ transforms in the following manner
\begin{eqnarray*}
x&\mapsto& x+vt+a\\
t&\mapsto& t+b
\end{eqnarray*}
The group element $g = (b,a,v)$ can be faithfully represented by a $3\times 3$ upper
triangular matrix,
\begin{equation}
  g = \begin{pmatrix} 1 & b & a \\ 0 & 1 & v \\ 0 & 0 & 1 \end{pmatrix}\; ,
\label{galgrpelem}
\end{equation}
so that matrix multiplication captures the group composition law.  This group, also known as the {\em Heisenberg group} in the mathematical and signal analysis literature, is a
central extension of the group of translations of $\mathbb R^2$, (translations in time and velocity.) The exponent giving this extension is
\begin{equation}
\xi_{\hbox{\tiny{H}}} (\mathbf x, \mathbf x') = bv'\; ,
\label{heis-grp-mult}
\end{equation}
where, $\mathbf x = (b, v), \; \mathbf x' = (b', v')$. In the physical literature one usually works with another extension of $\mathbb R^2$, the resulting group being  referred to as the {\em Weyl-Heisenberg group\/}. This latter group is constructed using  an exponent which is projectively equivalent to (\ref{heis-grp-mult}). We shall come back to this point later.

Next we construct a different kind of an extension of  the Galilei group $\mathcal G_0$, by
forming its semidirect product with $\mathcal D_2$, the two-dimensional dilation group, i.e., we introduce two dilations (of space and time). The resulting group $\mathcal G_0\rtimes \mathcal D_2$ will be denoted  $\mathcal{G}_{\hbox{\tiny{aff}}}$. If  the space and time dilations are given by $\sigma$ and $\tau$, respectively, and  a generic group element of $\mathcal{G}_{\hbox{\tiny{aff}}}$ is written $(b,a,v,\sigma,\tau)$, then the corresponding group composition law reads
\begin{eqnarray}
\lefteqn{(b,a,v,\sigma,\tau)(b^{\prime},a^{\prime},v^{\prime},\sigma^{\prime},
\tau^{\prime})}\nonumber\\
&&=(b+e^{\tau}b^{\prime},a+e^{\tau}b^{\prime}v+e^{\sigma}a^{\prime}, v+e^{\sigma-\tau}v^{\prime},\sigma+\sigma^{\prime},\tau+\tau^{\prime})\; .
\label{affgrp-law}
\end{eqnarray}
We shall refer to $\mathcal{G}_{\hbox{\tiny{aff}}}$ as the {\em affine Galilei group\/.} It has the matrix representation
\begin{equation}
(b,a,v,\sigma,\tau)_{\hbox{{\tiny{aff}}}}=\begin{bmatrix}e^{\sigma}&ve^{\tau}&a\\
0&e^{\tau}&b\\0&0&1\end{bmatrix}
\label{aff-gal-grp-mat}
\end{equation}

\section{From affine Galilei to extended Heisenberg, shearlet and wavelet groups}\label{sec:affgal-toheisen-etal}
In this section, starting from the affine Galilei group $\mathcal{G}_{\hbox{\tiny{aff}}}$, we first derive the family of
extensions $G^{p}_{\hbox{\tiny{H}}}$ of the Heisenberg group, originally obtained in
\cite{Taylor}. Following this, we shall  show how the reduced shearlet group, constructed in \cite{Shearlet} is in fact one of the above groups. Finally, we shall obtain the wavelet group as another subgroup of the affine Galilei group. 

In subsequent sections, using the matrix representations of two central extensions (one of them being a trivial extension) of the Galilei-Schr\"odinger group $\mathcal G_s$, we shall demonstrate  that the Weyl-Heisenberg group and the connected Stockwell group are subgroups of these centrally extended groups. In other words, we shall have shown that all the groups of interest in time-frequency analysis and signal processing are obtainable from a single group, the affine Galilei $\mathcal{G}_{\hbox{\tiny{aff}}}$.

\subsection{Extended Heisenberg group $G^{p}_{\hbox{\tiny{H}}}$ as  subgroup of affine Galilei group $\mathcal{G}_{\hbox{\tiny{aff}}}$}
 Let us construct a family of subgroups of the the affine Galilei group $\mathcal{G}_{\hbox{\tiny{aff}}} = \mathcal G_0\rtimes D_2$ by restricting the two dilations $\sigma$ and $\tau$ to lie on a line $\tau = m\sigma$, where $m$ is a constant.
 The special case where $m=2$ is called the Galilei-Schr\"odinger group \cite{mahara}. We shall come back to this group later.

  Consider first the the family of (non-isomorphic) extensions $G^{p}_{\hbox{\tiny{H}}}$ of the Heisenberg group, worked out in \cite{Taylor}. This family of groups  is parametrized by a real number $p$, where $-1<p\leq 1$. The corresponding group law reads
\begin{equation}
(b,a,v,\sigma)(b^{\prime},a^{\prime},v^{\prime},\sigma^{\prime})=(b+e^{\frac{\sigma}
{p+1}}b^{\prime},
a+e^{\sigma}a^{\prime}+e^{\frac{\sigma}{p+1}}vb^{\prime},
e^{\frac{p\sigma}{p+1}}v^{\prime}+v,
\sigma+\sigma^{\prime}).
\label{taylor-grp-law}
\end{equation}
The matrix representation of the above family of Lie groups, referred to in (\cite{Taylor}) as the {\em extended Heisenberg groups\/}, is easily seen to be
\begin{equation}
(b,a,v,\sigma)^{p}_{\hbox{\tiny{H}}}=\begin{bmatrix}e^{\sigma}&ve^{\frac{\sigma}{p+1}}
&a\\0&e^{\frac{\sigma}{p+1}}&b\\0&0&1\end{bmatrix}\; , \qquad -1<p\leq 1.
\label{taylor-grp-mat}
\end{equation}
Comparing with (\ref{aff-gal-grp-mat}),  we immediately see that the groups $G^{p}_{\hbox{\tiny{H}}}$ are subgroups of the (1+1) affine Galilei group $\mathcal{G}_{\hbox{\tiny{aff}}}$ of the type where the two dilations are restricted to the line $\tau = m\sigma$, with  $m=\frac{1}{p+1}$.

\subsection{Reduced shearlet group as  subgroup of the affine Galilei group $\mathcal{G}_{\hbox{{\tiny{aff}}}}$}\label{subsec:aff-gal-to-shear}
The reduced shearlet group $\mathbb S$, as described in \cite{Shearlet}, has a generic element,
\begin{equation*}
s=(\mu,\nu,\alpha,\beta),\;\;\mu\in\mathbb{R}^{+},\;\nu\in\mathbb{R}\;\hbox{and}\;
(\alpha,\beta)\in\mathbb{R}^{2},
\end{equation*}
with the multiplication law
\begin{eqnarray}
\lefteqn{(\mu_{1},\nu_{1},\alpha_{1},\beta_{1})(\mu_{2},\nu_{2},\alpha_{2},\beta_{2})}
\nonumber\\
&&=(\mu_{1}\mu_{2},\nu_{1}+\nu_{2}\sqrt{\mu_{1}},\alpha_{1}+
\mu_{1}\alpha_{2}+\nu_{1}\sqrt{\mu_{1}}\beta_{2},\beta_{1}+\sqrt{\mu_{1}}\beta_{2})\; .
\label{shear-mult-law}
\end{eqnarray}
The matrix representation for the group $\mathbb{S}$ is as follows
\begin{equation}
(\mu,\nu,\alpha,\beta)=\begin{bmatrix}\mu&\nu\sqrt{\mu}&\alpha\\0&\sqrt{\mu}&\beta\\
0&0&1\end{bmatrix}
\label{shear-mat}
\end{equation}
Comparing with  (\ref{taylor-grp-mat}), we see that this group corresponds to the special case $p=1$, i.e., $m = \dfrac 12$,
\begin{equation}
(b,a,v,\sigma)_{\mathbb{S}}:=(b,a,v,\sigma)^{p=1}_{\hbox{\tiny{H}}}=
\begin{bmatrix}e^{\sigma}&ve^{\frac{\sigma}{2}}&a\\0&e^{\frac{\sigma}{2}}&b\\0&0&1
\end{bmatrix},
\label{shearletmat}
\end{equation}
and the explicit  identification
\begin{eqnarray*}
e^{\sigma}&\longrightarrow&\mu\\
v&\longrightarrow&\nu\\
a&\longrightarrow&\alpha\\
b&\longrightarrow&\beta \; .
\end{eqnarray*}
Thus, the reduced shearlet group $\mathbb{S}$ is a member of the family of  extensions $G^{p}_{\hbox{\tiny{H}}}$ of Heisenberg group (with $p=1$) and hence also a subgroup of the $(1+1)$-affine Galilei group. $\mathcal{G}_{\hbox{\tiny{aff}}}$.

\subsection{Wavelet group as  subgroup of the affine Galilei group $\mathcal{G}_{\hbox{\tiny{aff}}}$}
\label{subsec:aff-gal-to-affine}
The connected affine group or the wavelet group is a two-parameter group $G^{\hbox{\tiny{aff}}}_{+}$ which consists of transformations on $\mathbb{R}$ given by
\begin{equation}
x\mapsto dx+t,
\label{aff-grf-action}
\end{equation}
where $x\in\mathbb{R}$, $d>0$ and $t\in\mathbb{R}$. Here $d$ and $t$ can be regarded as the dilation and translation parameters, respectively. The group law for this group is given by
\begin{equation}
(d_1,t_1)(d_2,t_2)=(d_{1}d_{2},d_{1}t_{2}+t_{1})
\label{aff-grp-law}
\end{equation}

The matrix representation of $G^{\hbox{\tiny{aff}}}_{+}$, compatible with the above group law, is given by
\begin{equation}
(d,t)=\begin{bmatrix}d&t\\0&1\end{bmatrix}
\label{aff-grp-mat}
\end{equation}

  In the matrix (\ref{shearletmat}) of the reduced shearlet group if we set $b=v=0$, we are left with
\begin{equation}
s\mid_{\hbox{\tiny{Wavelet}}}=\begin{bmatrix}e^{\sigma}&0&a\\0&e^{\frac{\sigma}{2}}&0\\
0&0&1\end{bmatrix}\; ,
\end{equation}
which is a $3\times 3$ faithful matrix representation of $G^{\hbox{\tiny{aff}}}_{+}$ with the following identification
\begin{eqnarray*}
d&\longrightarrow&e^{\sigma}\\
t&\longrightarrow&a\; ,
\end{eqnarray*}
i.e., we ave obtained the wavelet group as a subgroup of the reduced shearlet group $\mathbb{S}$ and hence of the affine Galilei group $\mathcal{G}_{\hbox{\tiny{aff}}}$.

Thus, so far we have obtained all the groups mentioned in Section \ref{sec:intro}, except for the Stockwell group, as subgroups of the affine Galilei group. Although we shall later obtain the Stockwell group as a subgroup of a trivial central extension of the Galilei-Schr\"odinger group, which is itself a subgroup of the affine Galilei group, we might mention already here that we could obtain the Stockwell group also as a trivial central extension of the wavelet group. In this sense, we could have started with a trivial extension of the affine Galilei group and obtained all the  groups mentioned in Section \ref{sec:intro} essentially as subgroups of it.

\section{Extensions of the affine Galilei and related groups}\label{sec:cent-extensions}

The Galilei group $\mathcal G_0$,  has a non-trivial central extension \cite{levy}, and in fact, there is only one such extension, up to projective equivalence. This extension, which we describe below, incorporates the quantum kinematics of a physical system in a space-time of
$(1+1)$-dimensions.

Let $M$ be a non-zero, positive real number; the
local exponent $\xi:\mathcal{G}_{0}\times\mathcal{G}_{0}\rightarrow \mathbb{R}$, giving the extension in question is:
\begin{equation}
\xi(g,g^{\prime})=M[va^{\prime}+\frac{1}{2}b^{\prime}v^{2}]\; ,
\label{galmult}
\end{equation}
where $g\equiv(b,a,v)$ and $g^{\prime}\equiv(b^{\prime},a^{\prime},v^{\prime})$ are elements of $\mathcal{G}_{0}$. We denote this extended group by $\mathcal G^M$; writing a
generic element of  $\mathcal{G}^{M}$  as $(\theta,b,a,v)$, the group multiplication law reads,
\begin{eqnarray}
\lefteqn{(\theta,b,a,v)(\theta^{\prime},b^{\prime},a^{\prime},v^{\prime})}\nonumber\\
&&=(\theta+\theta^{\prime}+M[va^{\prime}+\frac{1}{2}b^{\prime}v^{2}],
b+b^{\prime},a+a^{\prime}+vb^{\prime},v+v^{\prime})
\end{eqnarray}
We shall refer to $\mathcal G^M$ as the {\em quantum Galilei group\/.}

\subsection{Non-central extension of affine Galilei group}\label{subsec:non-cen-ext}
The group $\mathcal{G}_{\hbox{\tiny{aff}}}$ does not have non-trivial central
 extensions. Consequently, it cannot be used in quantum mechanics, since a trivial
 extension fails to generate mass \cite{mahara}. From a physical point of view, it is
 therefore more meaningful to take the quantum Galilei group $\mathcal G^M$ and to form its semidirect product with $\mathcal D_2$. This way, we arrive at $\mathcal G^M_{\hbox{\tiny{aff}}} = \mathcal G^M\rtimes \mathcal D_2$, which is a {\em non-central
 extension} of the affine Galilei group.  For simplicity we will call this group the {\em extended affine Galilei group\/}. Denoting  a generic group element of this group by $(\theta,b,a,v,\sigma,\tau)$, the group multiplication law reads
\begin{eqnarray}
\lefteqn{(\theta, b, a, v,\sigma,\tau)(\theta^\prime,b^\prime,a^\prime,v^\prime,\sigma^\prime,\tau^\prime)}
\nonumber\\
&&=(\theta+e^{2\sigma-\tau}\theta^{\prime}+M[e^{\sigma}va^{\prime}+\frac{1}{2}
e^{\tau}v^{2}b^{\prime}],b+e^{\tau}b^{\prime},a+e^{\tau}b^{\prime}v+e^{\sigma}a^{\prime}, v+e^{\sigma-\tau}v^{\prime},\nonumber\\
&&\hspace{3.3in}\sigma+\sigma^{\prime},\tau+\tau^{\prime})
\end{eqnarray}
The matrix representation of an element of $\mathcal{G}^{M}_{\hbox{\tiny{aff}}}$, consistent with the
above multiplication rule is
\begin{equation}
(\theta,b,a,v,\sigma,\tau)^{M}_{\hbox{\tiny{aff}}}=\begin{bmatrix}
e^{\sigma}&ve^{\tau}&0&a\\0&e^{\tau}&0&b\\Mve^{\sigma}&\frac{1}{2}Mv^{2}
e^{\tau}&e^{2\sigma-\tau}&\theta\\0&0&0&1\end{bmatrix}
\label{extaffgalrep}
\end{equation}

 As mentioned earlier (see \cite{levy}),  all the multipliers for the (1+1) dimensional quantum Galilei group $\mathcal G^M$ are equivalent, i.e., there is only one equivalence class in the multiplier group of the $(1+1)$-dimensional Galilei group $\mathcal{G}_{0}$. In other words $H^{2}(\mathcal{G}_{0},\mathbb{U}(1))$ is just one dimensional. It is noteworthy in this context that equation (\ref{extaffgalrep}) is a matrix representation of $\mathcal{G}^{M}_{\hbox{\tiny{aff}}}$ provided that the multiplier we choose, from the one dimensional group $H^{2}(\mathcal{G}_{0},\mathbb{U}(1))$ to obtain $\mathcal{G}^{M}$ during the two step construction of $\mathcal{G}^{M}_{\hbox{\tiny{aff}}}$, has the form $e^{i\xi(g_1,g_2)}$, with $\xi$ given by equation (\ref{galmult}).  Choosing another, though equivalent, multiplier will alter the form of the matrix  (\ref{extaffgalrep}).

\subsection{Galilei-Schr\"odinger group: central extensions}\label{subsec:galschrext}
 Let us consider the particular case of the  subgroup of $\mathcal{G}_{\hbox{\tiny{aff}}}$ when $\tau =2\sigma$, i.e., $m=2$ (or $p = -\dfrac 12$ in (\ref{taylor-grp-mat})). We denote the resulting one-dimensional dilation group by $\mathcal D_s$ and the corresponding subgroup of $\mathcal{G}_{\hbox{\tiny{aff}}}$ by $\mathcal G_s$, so that $\mathcal{G}_{s}=\mathcal{G}_{0}\rtimes\mathcal{D}_{s}$. In the literature, this group is known as {\em the Galilei-Schr\"odinger group} \cite{mahara}. It is easy to construct a central extension,
denoted $\mathcal{G}^{M}_{s}$, of $\mathcal{G}_{s}$ by $\mathbb{U}(1)$,  using a  local exponent $\xi:\mathcal{G}_{s}\times\mathcal{G}_{s}\rightarrow\mathbb{R}$, or equivalently, using the multiplier $\exp{i\xi}:\mathcal{G}_{s}\times\mathcal{G}_{s}\rightarrow \mathbb{U}(1)$. We mention in
this context that since we prefer working with addition rather than multiplication, we shall henceforth talk in terms of exponents rather than multipliers.

 We proceed to construct two extensions of the Galilei-Schr\"odinger group, using two equivalent multipliers, and a third extension using a trivial or exact multiplier. To do that we first note that the group multiplication law for $\mathcal{G}_{s}$ is given by
\begin{equation}
(b,a,v,\sigma)(b^{\prime},a^{\prime},v^{\prime},\sigma^{\prime})=(b+e^{2\sigma}b^{\prime},
a+e^{\sigma}a^{\prime}+e^{2\sigma}vb^{\prime},v+e^{-\sigma}v^{\prime},\sigma+\sigma^{\prime})
\end{equation}
where a generic element of the group is denoted as $(b,a,v,\sigma)$. Now using the exponent
\begin{equation}
\xi((b,a,v,\sigma);(b^{\prime},a^{\prime},v^{\prime},\sigma^{\prime}))=
M[ve^{\sigma}a^{\prime}+\frac{1}{2}v^{2}e^{2\sigma}b^{\prime}]\; ,
\label{gemesexpon}
\end{equation}
we obtain a central extension $\mathcal{G}^{M}_{s}$ of $\mathcal{G}_{s}$ by $\mathbb{U}(1)$.
The group law for the centrally extended group $\mathcal{G}^{M}_{s}$ therefore reads
\begin{eqnarray}
\lefteqn{(\theta,b,a,v,\sigma)(\theta^{\prime},b^{\prime},a^{\prime},v^{\prime},
\sigma^{\prime})}\nonumber\\
&&=(\theta+\theta^{\prime}+M[ve^{\sigma}a^{\prime}+\frac{1}{2}v^{2}e^{2\sigma}
b^{\prime}],b+e^{2\sigma}b^{\prime},a+e^{2\sigma}vb^{\prime}+e^{\sigma}
a^{\prime},v+e^{-\sigma}v^{\prime},\nonumber\\
&&\hspace{3.7in}\sigma+\sigma^{\prime})\;,
\label{gemesmultrule}
\end{eqnarray}
which is consistent with the matrix representation,
\begin{equation}
(\theta,b,a,v,\sigma)^{M}_{s}=\begin{bmatrix}e^{\sigma}&ve^{2\sigma}&0&a\\
0&e^{2\sigma}&0&b\\Mve^{\sigma}&\frac{1}{2}Mv^{2}e^{2\sigma}&1&\theta\\0&0&0&1\end{bmatrix}\; .
\label{gemesmatrix}
\end{equation}
Comparing (\ref{extaffgalrep}) and (\ref{gemesmatrix}) we easily see that $\mathcal{G}^{M}_{s}\subset\mathcal{G}^{M}_{\hbox{\tiny{aff}}}$, which is clear since we have just set $\tau = 2\sigma$. It ought to be noted here, that in going from $\mathcal G_0$ to $\mathcal{G}^{M}_{s}$, two extensions were involved: first we extended $\mathcal G_0$ to the Galilei-Schr\"odinger group $\mathcal G_s$, by taking the semidirect product of the former with the dilation group $\mathcal D_s$, and then doing a central extension of this enlarged group. We could equivalently have reversed the process, i.e., first done a central extension of $\mathcal G_0$ to obtain the quantum Galilei group $\mathcal G^M$ and then taken a semi-direct of this group with $\mathcal D_s$ to again arrive at $\mathcal{G}^{M}_{s}$. In other words, in this case the two procedures commute.

Next consider a second local exponent,  $\xi_1:\mathcal{G}_{s}\times\mathcal{G}_{s}\rightarrow\mathbb{R}$ given by
\begin{equation}
\xi_{1}((b,a,v,\sigma);(b^{\prime},a^{\prime},v^{\prime},\sigma^{\prime}))=
\frac{M}{2}[-vv^{\prime}b^{\prime}e^{\sigma}+va^{\prime}e^{\sigma}-av^{\prime}e^{-\sigma}]\; .
\label{secondexp}
\end{equation}
This exponent is easily seen to be equivalent equivalent to $\xi$, given in (\ref{gemesexpon}). Indeed, the difference of the above two exponents,
\begin{eqnarray}
\xi-\xi_{1}&=&\frac{M}{2}[v^{2}e^{2\sigma}b^{\prime}+vv^{\prime}b^{\prime}e^{\sigma}+
va^{\prime}e^{\sigma}+v^{\prime}ae^{-\sigma}]\nonumber\\
&=&\frac{M}{2}(a+e^{2\sigma}vb^{\prime}+e^{\sigma}a^{\prime})(v+e^{-\sigma}v^{\prime})-
\frac{M}{2}av-\frac{M}{2}a^{\prime}v^{\prime}
\label{diffexp}
\end{eqnarray}
is a trivial exponent. In other words (\ref{diffexp}) can be rewritten in terms of the continuous function $\zeta_{M}:\mathcal{G}_{s}\rightarrow\mathbb{R}$,
\begin{equation}
\xi-\xi_{1}=\zeta_{M}((b,a,v,\sigma)(b^{\prime},a^{\prime},v^{\prime},\sigma^{\prime}))-
\zeta_{M}(b,a,v,\sigma)-\zeta_{M}(b^{\prime},a^{\prime},v^{\prime},\sigma^{\prime}),
\end{equation}
where $\zeta_{M}(b,a,v,\sigma)=\frac{M}{2}av$.

Let $\mathcal{G}^{M\prime}_{s}$ denote the central extension of $\mathcal{G}_{s}$ by $\mathbb{U}(1)$ with respect to the exponent $\xi_1$ given by equation (\ref{secondexp}).  The group multiplication law for $\mathcal{G}^{M\prime}_{s}$ reads
\begin{eqnarray}
\lefteqn{(\theta,b,a,v,\sigma)(\theta^{\prime},b^{\prime},a^{\prime},v^{\prime},
\sigma^{\prime})}\nonumber\\
&&=(\theta+\theta^{\prime}+\frac{M}{2}[-vv^{\prime}b^{\prime}e^{\sigma}+
va^{\prime}e^{\sigma}-av^{\prime}e^{-\sigma}],b+e^{2\sigma}b^{\prime},
a+e^{\sigma}a^{\prime}+e^{2\sigma}vb^{\prime},\nonumber\\
&&\hspace{2.9in}v+e^{-\sigma}v^{\prime},\sigma+\sigma^{\prime})
\label{secondgrpmult}
\end{eqnarray}
The matrix representation for $\mathcal{G}^{M\prime}_{s}$, compatible with the group law, (\ref{secondgrpmult}) is
\begin{equation}
(\theta,b,a,v,\sigma)^{M\prime}_{s}=\begin{bmatrix}e^{\sigma}&-e^{-\sigma}b&0&a-vb\\
0&e^{-\sigma}&0&-v\\\frac{1}{2}Mve^{\sigma}&\frac{1}{2}Mae^{-\sigma}&1&\theta\\0&0&0&1
\end{bmatrix}\; .
\label{secondgr-mat}
\end{equation}

\bigskip

 Finally, we extend the Galilei-Schrodinger group $\mathcal{G}_{s}$ centrally by $\mathbb{U}(1)$ with respect to the trivial exponent $\xi_2:\mathcal{G}_{s}\times\mathcal{G}_{s}\rightarrow\mathbb{R}$ given by
\begin{equation}
\xi_{2}((b,a,v,\sigma);(b^{\prime},a^{\prime},v^{\prime},\sigma^{\prime}))
  =  ae^{-\sigma}(1-e^{-\sigma^{\prime}})-e^{\sigma-\sigma^{\prime}}
vb^{\prime}\; .
\label{trivialexp}
\end{equation}
We call this extension $\mathcal{G}^{T}_{s}$. Again, it is straight forward to verify the fact that the exponent given in (\ref{trivialexp}) is indeed trivial, since it can be rewritten in terms of the continuous function $\zeta_{T}:\mathcal{G}_{s}\rightarrow\mathbb{R}$,
\begin{eqnarray*}
\lefteqn{\xi_{2}((b,a,v,\sigma);(b^{\prime},a^{\prime},v^{\prime},\sigma^{\prime}))}\\
&&=\zeta_{T}(b,a,v,\sigma)+\zeta_{T}(b^{\prime},a^{\prime},v^{\prime},\sigma^{\prime})-
\zeta_{T}((b,a,v,\sigma)(b^{\prime},a^{\prime},v^{\prime},\sigma^{\prime}))\; ,
\end{eqnarray*}
where $\zeta_{T}(b,a,v,\sigma)=ae^{-\sigma}$. Thus, the group law for the trivially extended Galilei-Schrodinger group $\mathcal{G}^{T}_{s}$ reads
\begin{eqnarray}
\lefteqn{(\theta,b,a,v,\sigma)(\theta^{\prime},b^{\prime},a^{\prime},v^{\prime},
\sigma^{\prime})}\nonumber\\
&&=(\theta+\theta^{\prime}+[ae^{-\sigma}(1-e^{-\sigma^{\prime}})-
e^{\sigma-\sigma^{\prime}}vb^{\prime}],b+e^{2\sigma}b^{\prime},
a+e^{\sigma}a^{\prime}+e^{2\sigma}vb^{\prime},\nonumber\\
&&\hspace{2.8in}v+e^{-\sigma}v^{\prime},\sigma+\sigma^{\prime})
\end{eqnarray}
The matrix representation of $\mathcal{G}^{T}_{s}$ compatible with the above group law is given by
\begin{equation}
(\theta,b,a,v,\sigma)^{T}_{s}=\begin{bmatrix}1&ae^{-\sigma}&-e^{\sigma}v&\theta\\
0&e^{-\sigma}&0&1-e^{-\sigma}\\0&-e^{-\sigma}b&e^{\sigma}&e^{-\sigma}b\\0&0&0&1\end{bmatrix}
\label{triv-ext-mat}
\end{equation}

\section{From  Galilei-Schr\"odinger to Weyl-Heisenberg and  Stockwell groups}\label{sec:gal-schrod-toheisen-etal}
In this section we obtain the Weyl-Heisenberg and Stockwell groups as subgroups of the centrally extended Galilei-Schr\"odinger groups. We shall also re-derive the Heisenberg group, which by construction was a subgroup of the affine Galilei group $\mathcal{G}_{\hbox{\tiny{aff}}}$, this time as a subgroup of one of  the central extensions of the Galilei-Schr\"odinger group.

\subsection{Heisenberg  and  Weyl-Heisenberg groups as subgroups of centrally extended
Galilei-Schr\"odinger groups}\label{subsec:galschrd-to-heis-wh}
As mentioned in Section \ref{sec:aff-gal-grp}, the Heisenberg group is identical to  the $(1+1)$-Galilei group $\mathcal G_0$, which means that it is trivially a subgroup of the affine Galilei group $\mathcal{G}_{\hbox{\tiny{aff}}}$. Moreover, the Heisenberg group is a central extension of the two-dimensional translation group of the plane, via the local exponent $\xi_{\hbox{\tiny{H}}}$ in (\ref{heis-grp-mult}). As also indicated earlier, in the physical literature one uses a different, but projectively equivalent, exponent $\xi_{\hbox{\tiny{WH}}}$ (see (\ref{wh-loc-exp}) below) to do this extension, the resulting group being called the {\em Weyl-Heisenberg group\/.} Thus, although the Heisenberg and the Weyl-Heisenberg groups are projectively equivalent, we shall continue to differentiate between them in this paper. We now proceed to obtain these groups as subgroups of central extensions of the Galilei-Schr\"odinger group. Changing notations a bit let $(q,p)$ denote a point in the plane $\mathbb R^2$.

 In constructing the Heisenberg  group $G_{\hbox{\tiny{H}}}$ one uses the local exponent,
\begin{equation}
\xi_{\hbox{\tiny{H}}}((q,p);(q^{\prime},p^{\prime}))=pq^{\prime}\; .
\label{heisen-expon}
\end{equation}
Writing a general element of this group as
\begin{equation*}
g=(\theta,q,p),\;\;\theta\in\mathbb{R},\;\;(q,p)\in\mathbb{R}^{2},
\end{equation*}
the group multiplication law reads
\begin{equation}
(\theta,q,p)(\theta^{\prime},q^{\prime},p^{\prime})=(\theta+\theta^{\prime}+
pq^{\prime},q+q^{\prime},p+p^{\prime}),
\label{heisen-mult}
\end{equation}
with the matrix representation being
\begin{equation}
(\theta,q,p)_{\hbox{\tiny{H}}}=\begin{bmatrix}1&p&\theta\\0&1&q\\0&0&1\end{bmatrix}\; .
\label{heisen-matrix}
\end{equation}

Now we form the subgroup $\mathcal{G}^{M}_{s}\vert_{\hbox{\tiny{H}}}$ of the centrally extended Galilei-Schrodinger group $\mathcal{G}^{M}_{s}$ by setting  $b=\sigma=0$, $\theta\in\mathbb{R}$ and $(a,v)\in\mathbb{R}^{2}$. The matrix representation of $\mathcal{G}^{M}_{s}\vert_{\hbox{\tiny{H}}}$ then has the form (see  (\ref{gemesmatrix})):
\begin{equation}
(\theta,0,a,v,0)^{M}_{s}:=(\theta,a,v)^{M}_{s}\mid_{\hbox{\tiny{H}}}=
\begin{bmatrix}1&v&0&a\\0&1&0&0\\Mv&\frac{1}{2}Mv^{2}&1&\theta\\0&0&0&1\end{bmatrix}\; ,
\end{equation}
which under the identification
\begin{eqnarray}
Mv&\longrightarrow&p\nonumber\\
a&\longrightarrow&q\nonumber\\
\theta&\longrightarrow&\theta
\label{identific}
\end{eqnarray}
reduces to
\begin{equation}
(\theta,q,p)^{M}_{s}\mid_{\hbox{\tiny{H}}}=\begin{bmatrix}1&\frac{p}{M}&0&q\\0&1&0&0\\
p&\frac{p^2}{2M}&1&\theta\\0&0&0&1\end{bmatrix}\; .
\end{equation}
Here we assume that the mass term $M$ is never zero. The above $4\times 4$ matrix is a faithful representation of the Heisenberg group $G_{\hbox{\tiny{H}}}$, compatible with the group law (\ref{heisen-mult}).

Thus, the Heisenberg group constructed using the  $\xi_{\hbox{\tiny{H}}}$ in (\ref{heisen-expon}), can also be obtained as  a subgroup of the nontrivial central extension $\mathcal{G}^{M}_{s}$ of the Galilei-Schr\"odinger group.

To obtain the Weyl-Heisenberg group in a similar manner, consider the local exponent
\begin{equation}
\xi_{\hbox{\tiny{WH}}}((q,p);(q^{\prime},p^{\prime}))=\frac{1}{2}(pq^{\prime}-p^{\prime}q)\; .
\label{wh-loc-exp}
\end{equation}
It is straightforward to verify  that this exponent is equivalent to  $\xi_{\hbox{\tiny{H}}}$ in (\ref{heisen-expon}). Indeed,
\begin{eqnarray*}
\xi_{\hbox{\tiny{H}}}-\xi_{\hbox{\tiny{WH}}}&=&pq^{\prime}-\frac{1}{2}(pq^{\prime}-
p^{\prime}q)\\
&=&\frac{1}{2}pq^{\prime}+\frac{1}{2}p^{\prime}q\\
&=&\frac{1}{2}(p+p^{\prime})(q+q^{\prime})-\frac{1}{2}pq-\frac{1}{2}p^{\prime}q^{\prime}\\
&=&\zeta((q,p);(q^{\prime},p^{\prime}))-\zeta(q,p)-\zeta(q^{\prime},p^{\prime})\; ,
\end{eqnarray*}
where $\zeta$ is a real valued continuous function defined on the group of translations of $\mathbb{R}^{2}$,  and hence $\xi_{\hbox{\tiny{H}}}-\xi_{\hbox{\tiny{WH}}}$ is a trivial exponent. Using the  exponent $\xi_{\hbox{\tiny{WH}}}$ we extend the group of translations of $\mathbb{R}^{2}$ to form the Weyl-Heisenberg group $G_{\hbox{\tiny{WH}}}$, which then obeys the following group law:
\begin{equation}
(\theta,q,p)(\theta^{\prime},q^{\prime},p^{\prime})=(\theta+\theta^{\prime}+\frac{1}{2}
(pq^{\prime}-p^{\prime}q),q+q^{\prime},p+p^{\prime})
\label{weyl-heisen-mult}
\end{equation}
The matrix representation compatible with the above group law can be written as
\begin{equation}
(\theta,q,p)_{\hbox{\tiny{WH}}}=\begin{bmatrix}1&0&0&q\\0&1&0&-p\\
\frac{1}{2}p&\frac{1}{2}q&1&\theta\\0&0&0&1\end{bmatrix}\; .
\label{weyl-heisen-matrix}
\end{equation}
Forming now the subgroup $\mathcal{G}^{M\prime}_{s}\vert_{\hbox{\tiny{WH}}}$ of the centrally extended Galilei-Schr\"odinger group $\mathcal{G}^{M\prime}_{s}$, obtained by setting $b=\sigma=0$, $\theta\in\mathbb{R}$ and $(a,v)\in\mathbb{R}^{2}$ (see (\ref{secondgr-mat})), we get for its matrix
representation
\begin{equation}
(\theta,0,a,v,0)^{M\prime}_{s}:=(\theta,a,v)^{M\prime}_{s}\mid\!\!\!\;_{\hbox{\tiny{WH}}}=
\begin{bmatrix}1&0&0&a\\0&1&0&-v\\\frac{1}{2}Mv&\frac{1}{2}Ma&1&\theta\\0&0&0&1\end{bmatrix}\; .
\end{equation}
Making again the identification (\ref{identific}), this becomes
\begin{equation}
(\theta,q,p)^{M\prime}_{s}\mid\!\!\!\;_{\hbox{\tiny{WH}}}=
\begin{bmatrix}1&0&0&q\\0&1&0&-\frac{p}{M}\\\frac{1}{2}p&\frac{1}{2}Mq&1&\theta\\0&0&0&1
\end{bmatrix}\; .
\end{equation}
Here we assume once more that the mass term $M$ is not zero. While the above matrix is not exactly of the same form as the one given in (\ref{weyl-heisen-matrix}), it does reproduce the group multiplication rule (\ref{weyl-heisen-mult}). Moreover, the two matrix representations are  equivalent, via the intertwining matrix
\begin{equation*}
S=\begin{bmatrix}1&0&0&0\\0&\frac{1}{M}&0&0\\0&0&1&0\\0&0&0&1\end{bmatrix}\; ,
\end{equation*}
i.e., we have
\begin{equation*}
S\;(\theta,q,p)_{\hbox{\tiny{WH}}}S^{-1}=(\theta,q,p)^{M\prime}_{s}\mid_{\hbox{\tiny{WH}}}.
\end{equation*}
In this way we have shown that the Weyl-Heisenberg group $G_{\hbox{\tiny{WH}}}$ is a subgroup of the nontrivial central extension $\mathcal{G}^{M\prime}_{s}$ of Galilei-Schr\"odinger group.

\subsection{Connected Stockwell group as subgroup of the trivial central extension $\mathcal{G}^{T}_{s}$ of the Galilei-Schr\'odinger group}\label{subsec:stockwell-group}
The connected Stockwell group $G_{\hbox{\tiny{SW}}}$ (see \cite{stockwell1,stockwell} for definition and properties) can  be seen as a trivial central extension of a group $G^\prime_{\hbox{\tiny{aff}}}$, isomorphic to the connected affine group $G^{\hbox{\tiny{aff}}}_+$ (see (\ref{aff-grp-mat})). Given a group element $(\gamma,\delta)\in \mathbb{R}^{>0}\times\mathbb{R}$, we define the group law for $G^\prime_{\hbox{\tiny{aff}}}$  by
\begin{equation}
(\gamma_{1},\delta_{1})(\gamma_{2},\delta_{2})=(\gamma_{1}\gamma_{2},\delta_{1}+\frac{1}
{\gamma_{1}}\delta_{2})
\end{equation}
 Comparing with (\ref{aff-grp-law}), we identify the group homomorphism $f:G^{\hbox{\tiny{aff}}}_+ \longrightarrow G^\prime_{\hbox{\tiny{aff}}}$
\begin{equation}
f(\gamma,\delta)=(\frac{1}{\gamma},\delta)\; .
\end{equation}
Let us extend the   group $G^\prime_{\hbox{\tiny{aff}}}$ centrally using the exponent
\begin{eqnarray}
\xi_{s}((\gamma_{1},\delta_{1});(\gamma_{2},\delta_{2}))&=&\gamma_{1}\delta_{1}
(1-\gamma_{2})\nonumber\\
&=&\gamma_{1}\delta_{1}+\gamma_{2}\delta_{2}-(\gamma_{1}\gamma_{2})(\delta_{1}+
\frac{\delta_{2}}{\gamma_{1}})\; .
\label{stockwell-exp}
\end{eqnarray}
This is in fact a trivial exponent since it can be written in terms  of the continuous function $\zeta_{s}:G^\prime_{\hbox{\tiny{aff}}}\rightarrow\mathbb{R}$:
\begin{equation}
\xi_{s}((\gamma_{1},\delta_{1});(\gamma_{2},\delta_{2}))=\zeta_{s}(\gamma_{1},
\delta_{1})+\zeta_{s}(\gamma_{2},\delta_{2})-\zeta_{s}((\gamma_{1},\delta_{1})
(\gamma_{2},\delta_{2}))\; ,
\end{equation}
where $\zeta_{s}(\gamma,\delta)=\gamma\delta$.
The group so extended obeys the multiplication rule
\begin{equation}
(\theta_{1},\gamma_{1},\delta_{1})(\theta_{2},\gamma_{2},\delta_{2})=(\theta_{1}+\theta_{2}+
[\gamma_{1}\delta_{1}(1-\gamma_{2})],\gamma_{1}\gamma_{2},\delta_{1}+\frac{1}{\gamma_{1}}
\delta_{2})\; ,
\end{equation}
which is the product rule for elements of the Stockwell group $G_{\hbox{\tiny{SW}}}$. This proves that the Stockwell group is a trivial central extension of the wavelet or affine group. The matrix representation of a group element of $G_{\hbox{\tiny{SW}}}$ is seen to be
\begin{equation}
(\theta,\gamma,\delta)_{\hbox{\tiny{SW}}}=\begin{bmatrix}1&\gamma\delta&\theta\\
0&\gamma&1-\gamma\\0&0&1\end{bmatrix}\; .
\label{stockwell-mat}
\end{equation}

We now show that this group can also be obtained as a subgroup of the trivially extended
Galilei-Schr\"odinger group $\mathcal{G}_{s}^{T}$ (see ((\ref{trivialexp}) - (\ref{triv-ext-mat})). Indeed, comparing (\ref{trivialexp}) to (\ref{stockwell-exp}) it is clear that the former exponent reduces to he latter if $v$ is set equal to zero. Next, setting $v = b = 0$ in $\mathcal{G}_{s}^{T}$ we
see that (\ref{triv-ext-mat}) reduces to
\begin{equation}
(\theta,0,a,0,\sigma)^{T}_{\hbox{\tiny{s}}}:=
(\theta,a,\sigma)^{T}_{\hbox{\tiny{s}}}\mid_{\hbox{\tiny{SW}}}=
\begin{bmatrix}1&ae^{-\sigma}&0&\theta\\0&e^{-\sigma}&0&1-e^{-\sigma}\\0&0&e^{\sigma}&0\\
0&0&0&1\end{bmatrix}\; .
\end{equation}
The identification
\begin{eqnarray*}
e^{-\sigma}\longrightarrow \gamma\\
a\longrightarrow \delta\\
\theta\longrightarrow \theta
\end{eqnarray*}
and subsequent elimination of the redundant third row and column is then seen to yield the matrix  (\ref{stockwell-mat}).

\bigskip

We can conveniently depict all these various extensions and reductions to subgroups by means of a diagram.

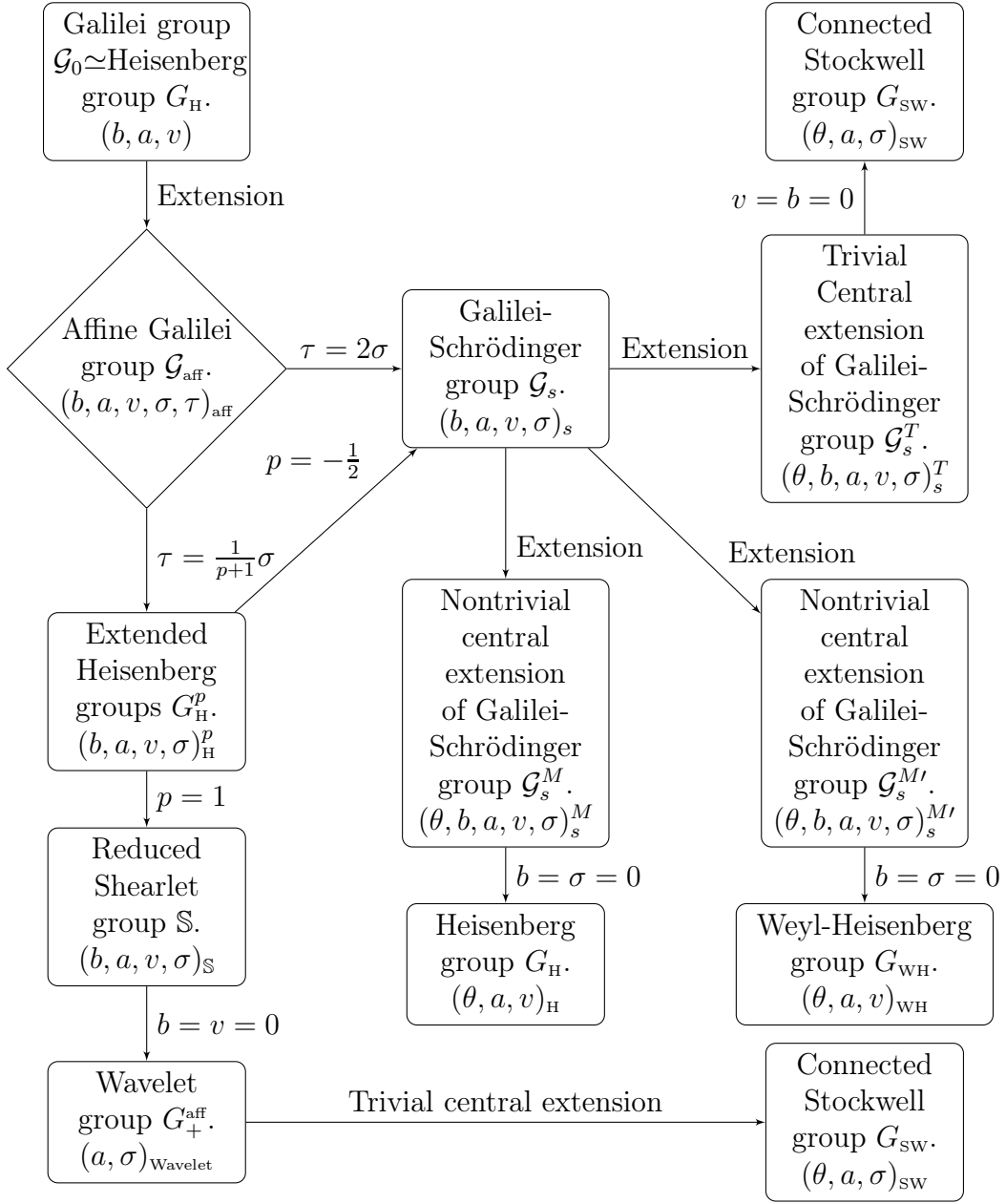
\begin{figure}
\tikzstyle{decision} = [diamond, draw,
    text width=6em, text badly centered, node distance=3cm, inner sep=0pt]
\tikzstyle{block} = [rectangle, draw,
    text width=6.3em, text centered, rounded corners, minimum height=4em]
\tikzstyle{block1} = [rectangle, draw,
    text width=6em, text centered, rounded corners, minimum height=4em]
\tikzstyle{block2} = [rectangle, draw,
    text width=8em, text centered, rounded corners, minimum height=3em]
\tikzstyle{line} = [draw, -latex']

\begin{tikzpicture}[node distance = 5cm, auto]
    \node [block](Zinit) {Galilei group $\mathcal{G}_{0}$$\simeq$Heisenberg group $G_{\hbox{\tiny{H}}}$.\\$(b,a,v)$};
    \node [decision, below of=Zinit, node distance=4cm] (init) {Affine Galilei group $\mathcal{G}_{\hbox{\tiny{aff}}}$.\\$(b,a,v,\sigma,\tau)_{\hbox{\tiny{aff}}}$};
    \node [block, right of=init, node distance=5cm] (ginit) {Galilei- Schr\"odinger group $\mathcal{G}_{s}$.\\$(b,a,v,\sigma)_{s}$};
    \node [block, below of=ginit, node distance=4.8cm] (Minit) {Nontrivial central extension of Galilei-Schr\"odinger group $\mathcal{G}^{M}_{s}$.\\$(\theta,b,a,v,\sigma)^{M}_{s}$};
    \node [block, right of=Minit, node distance=5cm](Linit){Nontrivial central extension of Galilei-Schr\"odinger group $\mathcal{G}^{M\prime}_{s}$.\\$(\theta,b,a,v,\sigma)^{M\prime}_{s}$};
    \node [block2, below of=Linit, node distance=3.5cm](Tinit){Weyl-Heisenberg group $G_{\hbox{\tiny{WH}}}$.\\$(\theta,a,v)_{\hbox{\tiny{WH}}}$};
    \node [block, right of=ginit, node distance=5cm] (Kinit) {Trivial Central extension of Galilei-Schr\"odinger group $\mathcal{G}^{T}_{s}$.\\$(\theta,b,a,v,\sigma)^{T}_{s}$};
    \node [block1, below of=Minit, node distance=3.5cm] (Winit) {Heisenberg group $G_{\hbox{\tiny{H}}}$.\\$(\theta,a,v)_{\hbox{\tiny{H}}}$};
    \node [block1, above of=Kinit, node distance=4cm] (Sinit) {Connected Stockwell group $G_{\hbox{\tiny{SW}}}$.\\$(\theta,a,\sigma)_{\hbox{\tiny{SW}}}$};
    \node [block1, below of=init, node distance=4.5cm] (Pinit) {Extended Heisenberg groups $G^{p}_{\hbox{\tiny{H}}}$.\\$(b,a,v,\sigma)^{p}_{\hbox{\tiny{H}}}$};
    \node [block1, below of=Pinit, node distance=3cm] (identify) {Reduced Shearlet group $\mathbb{S}$.\\$(b,a,v,\sigma)_{\mathbb{S}}$};
    \node [block1, below of=identify, node distance=3cm] (evaluate) {Wavelet group $G^{\hbox{\tiny{aff}}}_{+}$.\\$(a,\sigma)_{\hbox{\tiny{Wavelet}}}$};
    \node [block1, right of=evaluate, node distance=10cm] (Sinit1) {Connected Stockwell group $G_{\hbox{\tiny{SW}}}$.\\$(\theta,a,\sigma)_{\hbox{\tiny{SW}}}$};

    \path [line] (Zinit) --node {Extension} (init);
    \path [line] (init)--node {$\tau=2\sigma$} (ginit);
    \path [line] (ginit)--node [near end] {Extension} (Minit);
    \path [line] (ginit)--node {Extension} (Kinit);
    \path [line] (ginit)--node [near end]{Extension} (Linit);
    \path [line] (Linit)--node {$b=\sigma=0$} (Tinit);
    \path [line] (Minit)--node {$b=\sigma=0$} (Winit);
    \path [line] (init)--node {$\tau={\frac{1}{p+1}}\sigma$} (Pinit);
    \path [line] (Pinit)--node {$p=1$} (identify);
    \path [line] (Pinit)--node [near end] {$p=-\frac{1}{2}$} (ginit);
    \path [line] (Kinit)--node {$v=b=0$} (Sinit);
    \path [line] (identify) --node {$b=v=0$} (evaluate);
    \path [line] (evaluate) --node {Trivial central extension} (Sinit1);

\end{tikzpicture}
\caption{Flowchart showing the passage  from the (1+1)-affine Galilei group to the various groups of signal analysis.}
\end{figure}
\newpage
\section{Decomposition of UIRs of the affine Galilei group and central extensions of the Galilei-Schr\"odinger group restricted to various subgroups}\label{sec:UIR-decomp}
The general procedure for building signal transforms, starting from a group $G$ is first to define functions over the group using matrix elements of unitary irreducible representations. Provided these functions possess certain desirable properties which, among others, enable one to reconstruct the signal, they can be used as transforms describing the signal. In other words, the signal transforms are functions which encode the properties of the signal in terms of the group parameters. It is therefore of interest to construct unitary irreducible representations of the various groups discussed in the previous sections and to see how representations of the smaller subgroups, relevant to signal analysis, sit inside representations of the bigger groups.  

 The affine Galilei group $\mathcal{G}_{\hbox{\tiny{aff}}}$ was defined in Section \ref{sec:aff-gal-grp}, following which in Section \ref{sec:affgal-toheisen-etal} we studied its restriction to various subgroups of interest. In this section we shall first construct unitary irreducible representations of the affine Galilei group  and then study their restrictions to the reduced shearlet and wavelet  subgroups.

 In  later subsections we will find the UIRs of the two central extensions of the Galilei-Schr\"odinger and look at their restrictions to the Heisenberg group $G_{\hbox{\tiny{H}}}$ and the connected Stockwell group $G_{\hbox{\tiny{SW}}}$.

\subsection{UIRs of affine Galilei group restricted to the reduced shearlet group}
The group law and matrix representation of the  affine Galilei group $\mathcal{G}_{\hbox{\tiny{aff}}}$ was given in (\ref{affgrp-law}) and (\ref{aff-gal-grp-mat}).
From the matrix representation, we easily infer the semidirect product structure, $\mathcal{G}_{\hbox{\tiny{aff}}}=\mathcal{T}\rtimes\mathcal{V}$, where  $\mathcal{T}$ is an abelian subgroup, with generic element $(b,a)$ and  $\mathcal{V}$ is the subgroup generated by the elements $(v,\sigma,\tau)$. Now, the action of $(v,\sigma,\tau)$ on the element $(b,a)$ as determined by (\ref{affgrp-law}) is seen to be
\begin{equation}
(v,\sigma,\tau)(b,a)=(e^{\tau}b,e^{\tau}vb+e^{\sigma}a)
\end{equation}
We also have
\begin{equation}
(v,\sigma,\tau)^{-1}(b,a)=(e^{-\tau}b,e^{-\sigma}(a-vb))\; .
\label{semidir-prod-act}
\end{equation}
Now let $(E,p)$ denote a generic element of $\mathcal{T}^{*}$, the dual of $\mathcal{T}$ and the corresponding character by
\begin{equation*}
<(E,p)\mid(b,a)>=e^{i(Eb+pa)}
\end{equation*}
The action of $(v,\sigma,\tau)\in\mathcal{V}$ on $(E,p)\in\mathcal{T}^{*}$ is then defined by
\begin{eqnarray}
\lefteqn{<(v,\sigma,\tau)(E,p)\mid(b,a)>}\nonumber\\
&&=<(E,p)\mid(v,\sigma,\tau)^{-1}(b,a)>\nonumber\\
&&=<(E,p)\mid(e^{-\tau}b,e^{-\sigma}(a-vb))>\nonumber\\
&&=e^{i[(e^{-\tau}E-e^{-\sigma}pv)b+e^{-\sigma}pa]}\;,
\label{aff-grp-dual-action}
\end{eqnarray}
from  which we  easily find the dual action  $(E,p) \longrightarrow (\bar{E},\bar{p})$,
\begin{eqnarray}
\bar{E}&=&e^{-\tau}E-e^{-\sigma}pv\nonumber\\
\bar{p}&=&e^{-\sigma}p
\label{aff-grp-dual-action2}
\end{eqnarray}
which we can now use to compute the dual orbits.
We see that the sign of $p$ is an invariant for the same orbit while $E$ takes on all real values independently. In other words, the orbits are $(i)$ the two open half planes $\mathbb{R}\times \mathbb R^{\gtrless\; 0}$, one corresponding to all the positive values of $p$ and the other corresponding to the negative ones, $(ii)$ the two half lines $\mathbb R^{\gtrless \;0}$,  with $p = 0, \; E\gtrless 0$, and $(iii)$ the degenerate orbit $E=p=0$.
Note that none of these orbits are open-free (in the sense of \cite{Taylor2}).
Now using (\ref{semidir-prod-act}) and (\ref{aff-grp-dual-action2}) we obtain
\begin{equation}
(v,\sigma,\tau)^{-1}(E,p)=(E^{\prime},p^{\prime}) = (e^{\tau}(E+pv), e^{\sigma}p)
\label{dual-coords}
\end{equation}
From this it follows that
\begin{equation}
dE^{\prime}\; dp^{\prime} =e^{\sigma+\tau}\;dE\;dp\; , \qquad \text{on} \qquad \mathbb{R}\times \mathbb R^{\gtrless 0}\; ,
\label{measures1}
\end{equation}
and
\begin{equation}
dE^{\prime}=e^\tau\;dE\; , \qquad \text{on} \qquad \mathbb R^{\gtrless 0}\; .
\label{measures2}
\end{equation}

Using the Mackey's theory of induced representations \cite{Mackey,Projective}, we obtain four unitary irreducible representations of $\mathcal{G}_{\hbox{\tiny{aff}}}$, corresponding to the above four orbits. We denote the representations corresponding to the two half-planar orbits $\mathbb{R}\times \mathbb R^{\gtrless\; 0}$ by  $U^{\pm}_{\hbox{\tiny{aff}}}$, defined on $L^{2}(\mathbb{R}\times\mathbb{R}^{\pm}, dE\;dp)$, and the representations  on the half lines  $\mathbb R^{\gtrless\; 0}$, on $L^2 (\mathbb R^{\pm}, dE)$, by $V^{\pm}_{\hbox{\tiny{aff}}}$. The representations are easily computed to be
\begin{equation}
(U^{\pm}_{\hbox{\tiny{aff}}}(b,a,v,\sigma,\tau)\hat{\psi})(E,p)
 = e^{\frac{\sigma+\tau}{2}}e^{i(Eb+pa)}\hat{\psi}(e^{\tau}(E+pv),e^{\sigma}p)\; , \quad
 p \gtrless 0\; ,
\label{first-reps}
\end{equation}
and
\begin{equation}
(V^{\pm}_{\hbox{\tiny{aff}}}(b,a,v,\sigma,\tau)\hat{\psi})(E)
=e^{\frac{\tau}{2}}e^{iEb}\hat{\psi}(e^{\tau}E)\; , \quad E \gtrless 0\; .
\label{first-reps2}
\end{equation}
Note that the last two representations are trivial on the subgroup of $\mathcal{G}_{\hbox{\tiny{aff}}}$ with $a=v=\sigma=0$, i.e., the affine or wavelet group defined by the two remaining parameters $b, \tau$, and in fact, constitute the two unitary irreducible representations of that group. As is well known, these two representations of the affine group are square integrable and give rise to wavelet transforms.

We saw in Section \ref{subsec:aff-gal-to-shear} that the (reduced) shearlet group $\mathbb S$ is the subgroup of $\mathcal{G}_{\hbox{\tiny{aff}}}$ corresponding to $\tau = \dfrac 12 \sigma$. Restricting $U^{\pm}_{\hbox{\tiny{aff}}}$ in (\ref{first-reps}) to this subgroup we get
 \begin{equation}
(U^{\pm}_{\hbox{\tiny{aff}}}\mid_{\mathbb{S}}(b,a,v,\sigma)\hat{\psi})(E,p)=
e^{\frac{3\sigma}{4}}e^{i(Eb+pa)}\hat{\psi}(e^{\frac{\sigma}{2}}(E+pv),e^{\sigma}p)\; , \quad p \gtrless 0\; .
\label{shear-red}
\end{equation}
A quick examination of (\ref{aff-grp-dual-action2}) shows that $\mathbb R \times \mathbb R^{\gtrless \; 0}$ are both open free orbits under the action of $\mathbb S$. Also, as representations of the (reduced) shearlet group the two representations (\ref{first-reps}) are irreducible and hence square-integrable. Indeed, these are the representations used to build the shearlet transforms.

\subsection{UIRs of affine Galilei group $\mathcal{G}_{\hbox{\tiny{aff}}}$ restricted to the wavelet group}
We saw in Section \ref{subsec:aff-gal-to-affine} that the wavelet or affine group $G^{\hbox{\tiny{aff}}}_{+}$ could be obtained from the shearlet group as the subgroup with $b = v = 0$, or directly from the affine galilei group $\mathcal{G}_{\hbox{\tiny{aff}}}$ as the subgroup with $b = v = \tau = 0$.

Setting $b=v=\tau=0$ in the representations $U^{\pm}_{\hbox{\tiny{aff}}}$ in (\ref{first-reps}) we obtain
\begin{equation}
(U^{\pm}_{\hbox{\tiny{aff}}}\mid_{\hbox{\tiny{Wavelet}}}(0,a,0,\sigma,0)\hat{\psi})
(E,p)=e^{\frac{\sigma}{2}}e^{ipa}\hat{\psi}(E,e^{\sigma}p)
\label{wavelet-grp-rep}
\end{equation}
as  representations of the wavelet group $G^{\hbox{\tiny{aff}}}_{+}$ on $L^{2}(\mathbb{R}\times\mathbb{R}^{\pm}, dE\;dp)$. However, these representations are not irreducible. Indeed, noting that
$$  L^{2}(\mathbb{R}\times\mathbb{R}^{\pm}, dE\;dp) \simeq  L^{2}(\mathbb{R}, dE)\otimes L^{2}(\mathbb{R}^{\pm}, dp),$$
the representations (\ref{wavelet-grp-rep}) are immediately seen to be of the form
\begin{equation}
U^{\pm}_{\hbox{\tiny{aff}}}\mid_{\hbox{\tiny{Wavelet}}} = I\otimes U^{\pm}_{\text{\tiny{Wavelet}}}\; , \label{wavelet-grp-rep2}
\end{equation}
where $I$ is the identity operator on $L^{2}(\mathbb{R}, dE)$ and $U^{\pm}_{\text{\tiny{Wavelet}}}$ are the two unitary irreducible representations of $G^{\hbox{\tiny{aff}}}_{+}$ on $L^{2}(\mathbb{R}^{\pm}, dp)$, given by
\begin{equation}
  (U^{\pm}_{\text{\tiny{Wavelet}}}(a, \sigma)\hat{\psi})(p) = e^{\frac{\sigma}{2}}e^{ipa}\hat{\psi}(e^{\sigma}p)\; .
\label{wavelet-grp-rep3}
\end{equation}
 A decomposition of (\ref{wavelet-grp-rep2}) into irreducibles is easily done. Indeed, let $\{\hat{\phi}_n\}_{n=0}^\infty$ be an orthonormal basis of $L^{2}(\mathbb{R}, dE)$ and $\mathfrak H_n$ the one-dimensional subspaces spanned by $\hat{\phi}_n , \; n=0,1,2, \ldots , \infty$, so that $L^{2}(\mathbb{R}, dE) = \oplus_{n=0}^\infty \mathfrak H_n$. It is then immediately clear that
\begin{equation}
 U^{\pm}_{\hbox{\tiny{aff}}}\mid_{\hbox{\tiny{Wavelet}}}(0,a,0,\sigma,0) = \oplus_{n=0}^\infty \; U^{\pm, \; n}_{\text{\tiny{Wavelet}}}(a, \sigma)\; ,
\label{wavelet-grp-rep4}
\end{equation}
where $U^{\pm, \; n}_{\text{\tiny{Wavelet}}}$ is an irreducible representation of $G^{\hbox{\tiny{aff}}}_{+}$ which is simply a direct product of the trivial representation of the wavelet group on $\mathfrak H_n$ with the irreducible representation $U^{\pm}_{\text{\tiny{Wavelet}}}$ on $L^{2}(\mathbb{R}^{\pm}, dp)$ given in (\ref{wavelet-grp-rep3}). This decomposition also implies, that the shearlet transform, when restricted to the parameters of the wavelet group, decomposes into an infinite sum of wavelet transforms.

\subsection{UIRs of centrally extended Galilei-Schr\"odinger group $\mathcal{G}^{M}_{s}$ restricted to the Heisenberg group $G_{\hbox{\tiny{H}}}$}

The group law for the centrally extended Galilei-Schr\"odinger group $\mathcal{G}^{M}_{s}$, formed using the exponent $\xi$ in (\ref{gemesexpon}), is given by (\ref{gemesmultrule}) and the corresponding matrix representation by (\ref{gemesmatrix}). From the matrix representation one can deduce the semidirect product structure $\mathcal{G}^{M}_{s}=\mathcal{T}\rtimes\mathcal{V}$ where $\mathcal{T}$ is an abelian subgroup with generic element $(\theta,b,a)$ and $\mathcal{V}$ a semi-simple group consisting of the elements $(v,\sigma)$. Note that  that $\mathcal{V}$ is just the affine or wavelet group which also has a semidirect product structure, since
\begin{equation*}
(v_1,\sigma_1)(v_2,\sigma_2)=(v_{1}+e^{-\sigma_{1}}v_{2},\sigma_{1}+\sigma_{2}).
\end{equation*}
Now let $(q,E,p)$ denote a generic element of $\mathcal{T}^{*}$, the dual of $\mathcal{T}$ and consider the character
\begin{equation*}
<(q,E,p)\mid(\theta,b,a)>=e^{i(q\theta+Eb+pa)}\; .
\end{equation*}
The action of the subgroup $\mathcal{V}$ on the abelian subgroup $\mathcal{T}$ follows from (\ref{gemesmultrule})
\begin{equation}
(v,\sigma)(\theta,b,a)=(\theta+M[ve^{\sigma}a+\frac{1}{2}e^{2\sigma}v^{2}b],be^{2\sigma},
e^{\sigma}a+e^{2\sigma}vb)\; .
\end{equation}

Now the action of $(v,\sigma)\in\mathcal{V}$ on $(q,E,p)\in\mathcal{T}^{*}$ is defined by
\begin{eqnarray}
\lefteqn{<(v,\sigma)(q,E,p)\mid(\theta,b,a)>}\nonumber\\
&&=<(q,E,p)\mid(v,\sigma)^{-1}(\theta,b,a)>\nonumber\\
&&=<(q,E,p)\mid(\theta+M[-va+\frac{1}{2}v^{2}b],e^{-2\sigma}b,e^{-\sigma}(a-vb))>\nonumber\\
&&=e^{i[q\theta+(e^{-2\sigma}E-e^{-\sigma}pv+\frac{1}{2}qMv^{2})b+(e^{-\sigma}p-qMv)a]}
\label{gal-schroed-dual-action1}
\end{eqnarray}
Thus dual orbit elements $(\bar{q},\bar{E},\bar{p})$ corresponding  to a fixed value of $(q,E,p)$ are given by
\begin{eqnarray}
\bar{q}&=&q\nonumber\\
\bar{E}&=&e^{-2\sigma}E-e^{-\sigma}pv+\frac{1}{2}qMv^{2}\nonumber\\
\bar{p}&=&e^{-\sigma}p-qMv\; ,
\label{gal-schr-dual-act}
\end{eqnarray}
so that,
\begin{equation}
\bar{E}-\frac{\bar{p}^{2}}{2\bar{q}M}=e^{-2\sigma}(E-\frac{p^2}{2qM})
\end{equation}
where we assume that $q \neq 0$. Since $q$ remains invariant under the transformation (\ref{gal-schr-dual-act}), we take $\bar{q}=q=\kappa$. We thus get two dual orbits, the interior and exterior of the parabola given by $E-\frac{p^2}{2\kappa M}=0$, lying on the two-dimensional plane determined by $q=\kappa$ in the $\bar{q}$-$\bar{E}$-$\bar{p}$ space. The parabola  $E-\frac{p^2}{2\kappa M}=0$ itself determines an orbit and there are additional orbits when $q = 0$. Here we shall only consider the first two orbits, i.e., the interior and exterior of the parabola, for each non-zero $\kappa \in \mathbb R$. Let us introduce the new variables
\begin{eqnarray}
p&=&k_{1}\nonumber\\
E-\frac{p^2}{2\kappa M}&=&k_{2}
\end{eqnarray}
Then, for fixed value of $q=\kappa$, the coordinates $(k_1,k_2)$ are easily seen to transform  as
\begin{eqnarray}
\bar{k}_{1}&=&e^{-\sigma}k_{1}-\kappa Mv\nonumber\\
\bar{k}_{2}&=&e^{-2\sigma}k_{2}
\label{gal-schr-dual-act2}
\end{eqnarray}
In these new coordinates,
\begin{equation*}
(v,\sigma)(q,k_1,k_2)=(q,e^{-\sigma}k_{1}-qMv,e^{-2\sigma}k_{2})\; ,
\end{equation*}
and
\begin{equation}
(v,\sigma)^{-1}(k_1,k_2)=(e^{\sigma}(k_{1}+\kappa Mv),e^{2\sigma}k_{2}) := (k_1^\prime , k_2^\prime)\;, 
\end{equation}
so that,
\begin{eqnarray*}
k_{1}^{\prime}&=&e^{\sigma}(k_{1}+\kappa Mv)\\
k_{2}^{\prime}&=&e^{2\sigma}k_{2}\; .
\end{eqnarray*}
Therefore we obtain
\begin{equation}
dk_{1}^{\prime}\;dk_{2}^{\prime}=e^{3\sigma}dk_{1}\;dk_{2}
\end{equation}
Using again the method of induced representations, we arrive at the two UIRs of $\mathcal{G}^{M}_{s}$ defined on either $L^{2}(\mathbb{R}\times\mathbb{R}^{\pm},dk_{1}\;dk_{2})$, for each non-zero value of $q=\kappa$,
\begin{equation}
(U_{\pm}^{\kappa}(\theta,b,a,v,\sigma)\hat{\psi})(k_{1},k_{2})=e^{\frac{3\sigma}{2}}
e^{i(\kappa \theta+k_{1}a+\{k_{2}+\frac{(k_1)^2}{2\kappa M}\}b)}\hat{\psi}(e^{\sigma}(k_{1}+\kappa Mv),e^{2\sigma}k_{2})\; .
\label{gal-schrod-uir}
\end{equation}

Let us now go back to the Heisenberg group $G_{\hbox{\tiny{H}}}$, as discussed in Section \ref{subsec:galschrd-to-heis-wh} and construct its unitary irreducible representations, following similar techniques. From the matrix representation in (\ref{heisen-matrix}) we infer the semidirect product structure,
\begin{equation*}
G_{\hbox{\tiny{H}}}=\mathcal{T}\rtimes\mathcal{A}
\end{equation*}
where $(\theta,q)$ constitute elements of the  abelian subgroup $\mathcal{T}$ and $p$ is an element of the subgroup $\mathcal{A}$. Now $p\in\mathcal{A}$ acts on $(\theta,q)\in\mathcal{T}$ in the following manner
\begin{equation}
p(\theta,q)=(\theta+pq,q)
\end{equation}

We now denote by $(s,t)$ a geneirc element of $\mathcal{T}^{*}$, the dual of the abelian subgroup $\mathcal{T}$. Let us take the character
\begin{equation*}
<(s,t)\mid(\theta,q)>=e^{i(s\theta+tq)};
\end{equation*}
then
\begin{eqnarray}
<p(s,t)\mid(\theta,q)>&=&<(\bar{s},\bar{t})\mid(\theta,q)>\nonumber\\
&=& e^{i(\bar{s}\theta+\bar{t}q)}\nonumber\\
&=&<(s,t)\mid p^{-1}(\theta,q)>\nonumber\\
&=&<(s,t)\mid(\theta-pq,q)>\nonumber\\
&=& e^{i[s\theta+(t-sp)q]}
\end{eqnarray}
For fixed  $(s,t)$ the coordinates of its orbit  orbits under the action of $\mathcal A$ are
\begin{eqnarray}
\bar{s}&=&s\nonumber\\
\bar{t}&=&t-sp
\end{eqnarray}
Thus, the dual orbits are a family of parallel straight lines, one for each value of $s$ and $dt$ is the invariant measure on the orbit. Once again, using Mackey's theory of induced representation we obtain the UIR, corresponding to each dual orbit, i.e., for each fixed value of $s$:
\begin{equation}
(U^s_{\hbox{\tiny{H}}}(\theta,q,p)\hat{\psi})(t)=e^{is\theta}
e^{itq}\hat{\psi}(t+sp)\;,
\label{weyl-heis-rep}
\end{equation}
on the Hilbert space $L^2(\mathbb R, \; dt)$.

Now the restriction of the UIR (\ref{gal-schrod-uir}) of the centrally extended Galilei-Schr\"odinger group $\mathcal{G}^{M}_{s}$ to the Heisenberg group $G_{\hbox{\tiny{H}}}$ is seen to be
\begin{equation}
(U_{\pm}^{\kappa}\mid_{\hbox{\tiny{H}}}(\theta,0,a,v,0)\hat{\psi})(k_{1},k_{2})=e^{i(\kappa \theta+k_{1}a)}\hat{\psi}(k_{1}+\kappa Mv,k_{2})
\end{equation}
Thus,
\begin{equation}
U_{\pm}^{\kappa}\mid_{\hbox{\tiny{H}}}=U^{\kappa}_{\hbox{\tiny{H}}}\otimes I_\pm
\end{equation}
where $U^{\kappa}_{\hbox{\tiny{H}}}$ is the unitary irreducible representation of the Heisenberg group on  $L^{2}(\mathbb{R},dk_{1})$ and $I_\pm$ are the identity operators on $L^{2}(\mathbb{R}^{\pm},dk_{2})$. Once again we can decompose this representation as an infinite direct sum of irreducibles,
$$U_{\pm}^{\kappa}\mid_{\hbox{\tiny{H}}}= \oplus_{n=0}^\infty U^{\pm,\; n}_\kappa\; .$$
 just as in (\ref{wavelet-grp-rep4}). Here each $U^{\pm,\; n}_\kappa$ is a copy of the UIR (\ref{weyl-heis-rep}) with $s=\kappa$ on the Hilbert space $L^{2}(\mathbb{R},dk_{1})$ times a trivial representation on a one dimensional subspace of $L^{2}(\mathbb R^\pm,\;dk_{2})$.

We also recall that in Section \ref{subsec:galschrd-to-heis-wh} we obtained the Weyl-Heisenberg group $G_{\hbox{\tiny{WH}}}$ as a subgroup of the centrally extended Galilei-Schr\"odinger group $\mathcal{G}^{M\prime}_{s}$. We could just as well have obtained similar representations  of $G_{\hbox{\tiny{WH}}}$ and their decomposition into irreducibles from the UIR's of $\mathcal{G}^{M\prime}_{s}$.

\subsection{UIRs of cenrally extended (trivial) Galilei-Schr\"odinger group $\mathcal{G}^{T}_{s}$ restricted to connected Stockwell group}\label{gts-rep-restric}
In Section \ref{subsec:galschrext} we had introduced the Galilei-Schr\"odinger group $\mathcal{G}_{s}$, by setting  $\tau=2\sigma$ in the affine Galilei group (see (\ref{aff-gal-grp-mat})). Later we obtained a central extension of it using the trivial exponent $\xi_2$ in (\ref{trivialexp}). Here we shall obtain UIRs of this centrally extended group by first finding  unitary irreducible representations of $\mathcal{G}_{s}$ itself. The  matrix representation of  $\mathcal{G}_{s}$ is  found by substituting $\tau=2\sigma$ in (\ref{aff-gal-grp-mat}):
\begin{equation}
(b,a,v,\sigma)_{s}=\begin{bmatrix}e^{\sigma}&ve^{2\sigma}&a\\0&e^{2\sigma}&b\\0&0&1
\end{bmatrix}\; .
\label{gal-schrod-mat}
\end{equation}
From this follows the semi-direct product structure, $\mathcal{G}_{s}=\mathcal{T}\rtimes\mathcal{V}$ where the abelian subgroup $\mathcal{T}$ consists of elements  $(b,a)$ and the subgroup $\mathcal{V}$ consists of the elements $(v,\sigma)$.

Now let $(E,p)$ denote a generic element of $\mathcal{T}^{*}$, the dual to $\mathcal{T}$ and consider the corresponding character
\begin{equation*}
<(E,p);(b,a)>=e^{i(Eb+pa)}\; .
\end{equation*}
The action of the subgroup $\mathcal{V}$ on the abelian subgroup $\mathcal{T}$ can be immediately read off. We find,
\begin{equation*}
(v,\sigma)^{-1}(b,a)=(e^{-2\sigma}b,e^{-\sigma}(a+vb))\;,
\end{equation*}
and the action of $(v,\sigma)\in\mathcal{V}$ on $(E,p)\in\mathcal{T}^{*}$:
\begin{equation*}
<(v,\sigma)(E,p);(b,a)>=e^{i[(e^{-2\sigma} E+e^{-\sigma}pv)b+e^{-\sigma}pa]}
\end{equation*}
Thus, writing
\begin{equation*}
(v,\sigma)^{-1}(E,p)=(E^{\prime},p^{\prime})
\end{equation*}
we get the equations for the dual orbit, corresponding to $(E, p)$
\begin{eqnarray}
E^{\prime}&=&e^{2\sigma}(E+pv)\nonumber\\
p^{\prime}&=&pe^{\sigma}
\end{eqnarray}
We shall only consider orbits for which $p\neq 0$. Making a change of variables $(E,p)\mapsto(t=\frac{E}{p^{2}},p)$, the orbit equations become
\begin{eqnarray}
t^{\prime}&=&t+\frac{v}{p}\nonumber\\
p^{\prime}&=&pe^{\sigma}
\label{orb-eqns}
\end{eqnarray}
Thus we get two orbits in the $t$-$p$ space, namely, the two disjoint open half planes ($p \gtrless 0$). Also,
\begin{equation}
dt^{\prime}\!\;dp^{\prime}=e^{\sigma}dt\!\;dp
\end{equation}

Again, following the standard Mackey construction we get the following two unitary irreducible representations of the ordinary Galilei-Schr\"odinger group, corresponding to these two orbits $\mathbb{R}\times\mathbb{R}^{\pm}$ in the  $t$-$p$ space:
\begin{equation}
(U^\pm(b,a,v,\sigma)\hat{\psi})(t,p)=e^{i(tp^{2}b+pa)}e^{\frac{\sigma}{2}}\hat{\psi}
(t+\frac{v}{p},e^{\sigma}p)
\end{equation}
The representations are carried by the Hilbert spaces $L^2 (\mathbb R \times \mathbb R^\pm , \; dt\; dp)$, respectively.

In  Section \ref{subsec:galschrext} the trivial exponent $\xi_2$ was shown to arise from the continuous function $\zeta_{T}:\mathcal{G}_{s}\rightarrow\mathbb{R}$ given by
\begin{equation}
\zeta_{T}(g)=ae^{-\sigma}\; .
\end{equation}
where $g\equiv(b,a,v,\sigma)$ is a generic element of $\mathcal{G}_{s}$. In terms of this continuous function it follows immediately that $\tilde{U}^\pm(g)=e^{i\zeta_{T}(g)}U^\pm(g)$ are projective representations of the Galilei-Schrodinger group $\mathcal{G}_{s}$. In other words, $U^{T, \pm}_{s}(\theta,b,a,v,\sigma):=e^{i\theta}\tilde{U}^\pm(b,a,v,\sigma)$ are unitary irreducible representations of the trivial central extension $\mathcal{G}^{T}_{s}$ of the Galilei-Schr\"odinger group.

 Next the UIRs $U^{T,\pm}_{s}$  restricted to the connected Stockwell group have the form
\begin{equation}
(U^{T,\pm}_{s}\mid_{\hbox{\tiny{SW}}}(\theta,0,a,0,\sigma))(t,p)=e^{i(\theta+ae^{-\sigma})}
e^{ipa}e^{\frac{\sigma}{2}}\hat{\psi}(t,e^{\sigma}p)
\label{stock-grp-subrep}
\end{equation}
Thus,
\begin{equation}
U^{T,\pm}_{s}\mid_{\hbox{\tiny{SW}}}=I\otimes U^{\pm}_{\hbox{\tiny{SW}}}
\label{res-stck-rep}
\end{equation}
where $I$ is the identity operator on $L^{2}(\mathbb{R},dt)$ and $U^{\pm}_{\hbox{\tiny{SW}}}$ are UIRs of the connected Stockwell group on $L^{2}(\mathbb{R}^\pm,\; dt)$. The representation (\ref{res-stck-rep}) again decomposes in the usual manner into an infinite direct sum of irreducibles.

We remark here that the UIRs of the Stockwell group $G_{\hbox{\tiny{SW}}}$ are not square-integrable (over the whole group). However, since taking $\theta = 0$ in (\ref{stock-grp-subrep}) yields a projective  representation of the affine group, the two non-trivial representaions  of which are both square-integrable, this fact can be exploited to arrive at square-integrability over the homogeneous space $G_{\hbox{\tiny{SW}}}/\Theta$, where $\Theta$ is the phase subgroup. This is exactly the sense in which square-integrability for representations of the Stockwell group has been defined in \cite{stockwell} and is in accordance with the theory of square-integrability modulo subgroups (see, for example \cite{coherent}).

\section{Conclusion}
The fact that the various groups of signal analysis  enumerated in Section \ref{sec:intro} are all obtainable from the affine Galilei group shows a remarkable unity in their structures and consequently of their unitary irreducible representations. In later publications we propose to make a comparative study of the structures of their co-adjoint orbits and Wigner functions built on them. From the point of view of signal transforms, all this could lead to a deeper understanding of how signal transforms, defined over a larger set of parameters, reduce when a smaller set of parameters is used, with the original signal still being reconstructible from the smaller set.


\end{document}